\documentstyle[aps,prb,epsf]{revtex}
\def\narrowtext{} \tighten \twocolumn

\newcommand{\bscco}{Bi$_2$Sr$_2$CaCu$_2$O$_{8+\delta}$}
\newcommand{\ybco}{YBa$_2$Cu$_3$O$_{7-y}$}

\begin{document}
\draft

\title{Where is the $\pi$ resonance?}
\author{O. Tchernyshyov,$^{1,2}$ M. R. Norman,$^2$
and A. V. Chubukov$^{2,3}$  }
\address{$^1$School of Natural Sciences, Institute for Advanced Study, 
Princeton, New Jersey, 08540}
\address{$^2$Materials Science Division, Argonne National Laboratory, 
Argonne, Illinois 60439}
\address{$^3$Department of Physics, University of Wisconsin,
Madison, Wisconsin 53706}

\date{September 5, 2000}

\twocolumn[\hsize\textwidth\columnwidth\hsize\csname 
  @twocolumnfalse\endcsname

\maketitle
\begin{abstract}
We discuss the interplay of particle-particle and particle-hole
spin-triplet channels in high-$T_c$ superconductors using a
quasiparticle dispersion motivated by angle-resolved photoemission.
Within a generalized RPA, we find a well defined antibound state of
two holes, the $\pi$ resonance of Demler and Zhang, as well as a bound
state of a particle and a hole, the spin exciton.  We show that the
energy of the $\pi$ resonance always exceeds $2\Delta$, twice the
maximum d-wave gap, therefore the neutron resonance observed in the
cuprates around energy $\Delta$ is most likely a spin exciton.  At the
same time, we speculate that the $\pi$ particle can exist at higher
energies and might be observed in neutron scattering around 100 meV.
\end{abstract}
]

\narrowtext

Reflecting our poor understanding of high-$T_c$ superconductivity in
general, theoretical debates continue on virtually every aspect of it.
A good example is the resonance observed in inelastic neutron
scattering at energies 25--43 meV in \ybco\/ and
\bscco.\cite{peak-discovery-ybco,peak-discovery-bscco} Measurements
using polarized neutrons\cite{neutrons-polarized} indicate that the
observed excitation is magnetic in nature.  Theoretical proposals
include a van Hove singularity in the Stoner
continuum,\cite{Lavagna,Abrikosov} a bound state (spin exciton) in
the particle-hole
channel,\cite{Levin,Mazin,Scalapino,Brinckmann,Abanov} and an
antibound state in the particle-particle channel ($\pi$
resonance).\cite{DZ,DKZ} Deciding between these approaches is somewhat
difficult since in all theories the resonance mode is an excited pair
of quasiparticles with total lattice momentum ${\bf Q}=(\pi/a,\pi/a)$
and spin $S=1$.\cite{Brinckmann,DKZ,Greiter}

In this paper, we use available experimental data and a simple
kinematic argument to demonstrate that a particle-hole bound state is
the most likely explanation for the resonance.  In a nutshell, our
reasoning goes as follows.  Angle-resolved photoemission spectroscopy
(ARPES) directly measures the dispersion of fermionic quasiparticles
$\epsilon_{\bf k}$ in the superconducting state.  From this, one can
deduce information about the continuum of states of two
quasiparticles with total momentum ${\bf Q}$ and, in particular,
determine its energy bounds.
An analysis of ARPES data in \bscco\/
shows that the two-quasiparticle continuum starts at $E_{\rm min}
\approx 2\Delta$, where $\Delta$ is the maximum value of the
superconducting gap at the
Fermi surface.  The commensurate neutron resonance in  
\bscco\/ resides at $E_{\rm res} \approx \Delta$, definitely below the
lower edge of the continuum.  The resonance  then can only be 
a bound state of
two quasiparticles, not an antibound state. 

\begin{figure}
\begin{center}
\leavevmode
\epsfxsize 2.9in
\epsffile{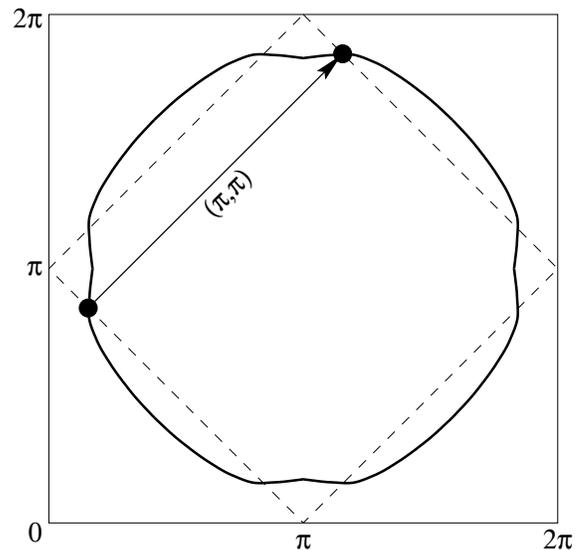}
\end{center}
\caption{``Hot spots'' are points on the Fermi surface connected by
the wave vector ${\bf Q}=(\pi/a,\pi/a)$ or those equivalent to it.
Equivalently, they are given by intersections of the Fermi surface with
the magnetic zone boundary (dashed lines).}
\label{Fig-hot-spots}
\end{figure}

This argument---that the lower edge of the two-particle continuum is
determined by $\Delta$ and not by the chemical potential $\mu$ as in
Ref.~\CITE{DZ}---is based on the analysis of the Fermi surface
inferred from ARPES data (Fig.~\ref{Fig-hot-spots}). We have verified
that, for this Fermi surface, $E_{\rm min}$ is determined not by
electrons along zone diagonals $k_x=k_y$, but rather by electrons near
the hot spots (points at ${\bf k=k}_F$ separated by ${\bf Q}$).  The
latter are located close to $(0,\pi)$ and symmetry related points, and
in the superconducting state have a gap $\Delta ({\bf k}_{\rm hs})
\approx \Delta$.  This yields a threshold at $E_{\rm min} \approx
2\Delta$.  Note that this argument also invalidates the interpretation
of the resonance as a van Hove singularity in the two-quasiparticle
continuum.\cite{Abrikosov}

Does it mean that the $\pi$ resonance does not exist?  Not
necessarily.  In our model calculations, the spin exciton and the
$\pi$ resonance are found to coexist at intermediate coupling
strengths, similar to earlier findings.\cite{DKZ} Furthermore, in a
situation where the continuum of states with two holes is narrow and
has a sharp upper edge, we find that the $\pi$ resonance is rather
sharp.  To which extent these conditions are satisfied in the cuprates
is not clear because self-energy effects could wash out the upper edge
of the continuum.  We therefore can only speculate that the $\pi$
resonance might be discovered in neutron scattering at energies above
$E_{\rm min}$ (60 meV in optimally doped \bscco), more likely around
100 meV.

To proceed, we need a model treating all spin-triplet excitations
(with charges 0 and $\pm2$) on an equal footing.  The simplest
approach is an adaptation of Anderson's treatment of phase
fluctuations and plasmons in a superconductor\cite{Anderson} to the
spin-triplet channel.  Precisely this method, known as generalized
RPA,\cite{Rickayzen,Bardasis} was used by Demler {\em et
al.}\cite{DKZ}.  To make the approach self-consistent, we make a
quasiparticle approximation with a dispersion\cite{Norman} fitted to
peak positions in the ARPES data in the superconducting state of
\bscco.  As the measured dispersion is flat near $(0,\pi)$, our
calculation yields a strikingly narrow two-electron continuum with a
width of about 10 meV.  This narrowness is caused by the proximity of
the hot spots to the van Hove points at $(\pi,0)$.

This quasiparticle model is a ``best case'' scenario for the antibound
state.  This model assumes that one has a superfluid Fermi liquid,
which appears to be consistent with transport and ARPES data, at least
at low temperatures.  Within this approximation, the upper edge of the
two-hole continuum is sharp, which is favorable for the formation of
the $\pi$ resonance.  By making such an approximation, we essentially
neglect the large incoherent part of the electron spectral function
known to exist from photoemission.  On the other hand, this incoherent
part is absent below the energy scale $\Delta + E_{\rm res}$ set by
the spin resonance,\cite{Abanov,ND} so the fermionic incoherence
affects the two-particle propagators only at energies above
$2\Delta+E_{\rm res}$.  Thus the $\pi$ resonance can, in principle,
live below this threshold.  We make no attempt to explain {\em why}
the quasiparticle dispersion is so remarkably flat near $(\pi,0)$.  A
number of authors\cite{ND,Dagotto,flat} have pointed out that this
flatness most likely results from a strongly $\omega$-dependent
self-energy which renormalizes the bare dispersion.  Also note that
the extreme narrowness of the two-particle continuum is not a crucial
part of our argument: we have observed similar behavior with a much
wider (70 meV) two-particle continuum based on normal state
dispersions.

We now proceed with the calculations.  Consider a system of fermions
with dispersion $\epsilon_{\bf k}$ given in Ref.~\CITE{Norman} and the
nearest-neighbor interaction~\cite{DZ,BA}
\begin{equation}
H_1 = \sum_{\langle ij \rangle}
J\left({\bf S}_i\!\cdot\!{\bf S}_j - \frac{1}{4}n_i n_j\right)
+ V n_i n_j.
\label{interaction}
\end{equation}
The $J$ term gives rise to an attraction in the particle-particle
d-wave singlet channel and in the particle-hole triplet channel, where it
can induce a spin exciton.  The $V$ term accounts for repulsion in the
triplet pairing channel and gives rise to the $\pi$ resonance.  In the
conventional $t$--$J$ model, $V = 0$. Following Refs.~\CITE{DZ} and
\CITE{BA}, we consider a more general interaction and treat $V$ and
$J$ as independent parameters.

Although we are interested in spin susceptibility, in a
superconducting ground state an operator of spin has the same quantum
numbers as an operator creating a spin-triplet pair.  We therefore
analyze the linear response for the set of three operators~\cite{DKZ}
\begin{eqnarray}
S_+ &=& N^{-1/2}\sum_{\bf k} 
a^\dagger_{{\bf k}\uparrow} a_{{\bf k+Q}\downarrow},
\label{S}\\
\pi &=& N^{-1/2}\sum_{\bf k} 
g_{\bf k}\, a_{{\bf Q}-{\bf k}\downarrow} a_{{\bf k}\downarrow}, 
\label{pi}\\
\bar\pi &=& N^{-1/2}\sum_{\bf k} 
g_{\bf k}\, a^\dagger_{{\bf Q}-{\bf k}\uparrow} a^\dagger_{{\bf k}\uparrow}.
\label{pi-bar}
\end{eqnarray}
Here $a_{{\bf k},\sigma}$ is an electron annihilation operator and
$g_{\bf k} = \cos{(k_x a)}-\cos{(k_y a)}$.  Operators $S_+$, $\pi$ and
$\bar\pi$ destroy a bosonic excitation with the same momentum ${\bf
Q}=(\pi/a,\pi/a)$ and spin $S_z=-1$, but with different charges $0$
and $\pm2$, respectively.  

\begin{figure}
\begin{center}
\leavevmode \epsfxsize 3.25in \epsffile{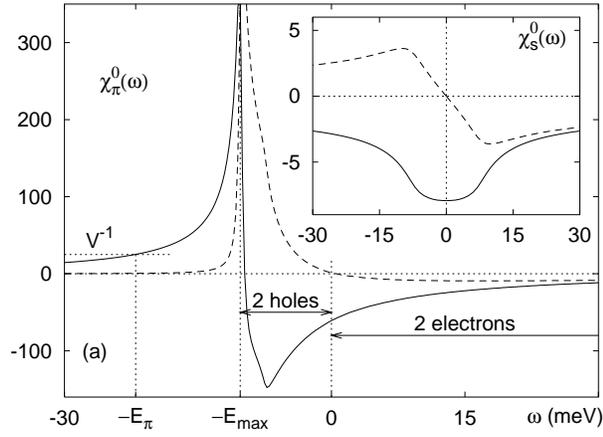}
\end{center}
\begin{center}
\leavevmode
\epsfxsize 3.25in
\epsffile{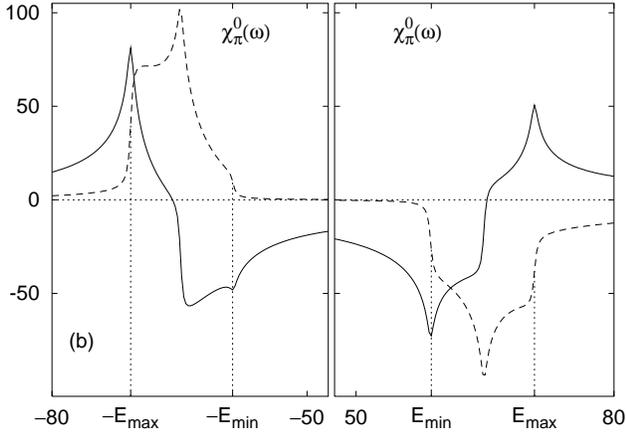}
\end{center}
\caption{The bare response function ($T=0$) of the operator $\pi$ in a
hypothetical Fermi liquid (a) and a d-wave superconductor with
$\Delta_{\rm max}$=35 meV (b) at ${\bf Q}$.  Solid line: real part, dashed
line: imaginary part.  Insert: spin susceptibility in the normal state
(note the difference in vertical scales).  A broadening of
$\Gamma$=0.5 meV was employed.  Solution of the equation
$\chi_\pi^0(\omega) = 1/V$ gives the RPA energy of the $\pi$
resonance, $\omega=-E_{\pi}$.  Note the difference in vertical
scales.  Horizontal arrows indicate two-particle continua.}
\label{Fig-bare-pair}
\end{figure}

As a warmup exercise, consider first a hypothetical (Fermi liquid)
non-superconducting ground state.  The triplet channels
(\ref{S}--\ref{pi-bar}) are decoupled by virtue of a charge U(1)
symmetry.  The Fourier transform of the bare pair susceptibility
\begin{equation}
\chi_{\pi}^0(t) = -i\theta(t)
\left.\langle[\pi(t),\pi^\dagger(0)]\rangle\right|_{V=J=0}
\label{def}
\end{equation}
is shown in Fig.~\ref{Fig-bare-pair}(a). As anticipated, the two-hole
continuum ($\omega<0$) is strikingly narrow: its upper edge is at
$|E_{\rm max}| \sim 10$ meV. The $\omega^{-1/2}$ divergence near
$E_{\rm max}$ is a van Hove-type singularity associated with the
$(\pi,0)$ points.  Observe that, at $|\omega|> |E_{\rm max}|$,
$\chi_{\pi}^0 (\omega)$ is purely real.  On the other hand, the bare
spin susceptibility $\chi_{S}^0 (\omega)$ [shown in the insert of
Fig.~\ref{Fig-bare-pair}(a)] is complex for all frequencies.

Within the generalized RPA approximation, the full susceptibilities
are given by
\begin{equation}
\chi_{\pi}(\omega) = \frac{\chi_{\pi}^0(\omega)}{1-V\chi_{\pi}^0(\omega)},
\hskip 3mm
\chi_{S}(\omega) = \frac{\chi_{S}^0(\omega)}{1+2J\chi_{S}^0(\omega)}.
\label{new}
\end{equation}
Note the opposite signs of the interaction terms: attractive in the
particle-hole channel, repulsive in the particle-particle channels.
Because $\chi_\pi^0(\omega)=1/V$ has a solution for any $V>0$
[Fig. ~\ref{Fig-bare-pair}(a)], the full $\pi$
susceptibility $\chi_{\pi}(\omega)$ acquires a pole above the upper
edge of the continuum.  This pole is the $\pi$ resonance.  In
contrast, no pole occurs in the RPA response for $S_+$.  This is a
consequence of the fact that the particle-hole continuum [and, hence,
${\rm Im}\chi_S (\omega)$] extends to the lowest energies.

We now turn to the superconducting state.  To facilitate the RPA
treatment, we assume that the superconducting state is of the BCS type
with a $d$-wave gap $\Delta({\bf k}) = \Delta(\cos{k_x a}-\cos{k_y
a})/2$. We treat $\Delta$ as another input parameter in the problem.
In principle, the gap has to be computed self-consistently,
 but for our purposes this is not necessary as we are not concerned with
response functions in the singlet channel at small momenta.

Since in a superconductor charge is defined modulo 2, all three
operators (\ref{S}-\ref{pi-bar}) now carry identical quantum numbers
and one can use any superposition of these.  It is customary to choose
\begin{equation}
A_0 = S_+,
\hskip 2mm 
A_1 = \frac{\pi-\bar\pi}
{\sqrt{2}},
\hskip 2mm 
A_2 = \frac{\pi+\bar\pi}{\sqrt{2}}.
\label{A_q}
\end{equation}
The operators $A_1$ and $A_2$ describe fluctuations of the phase and
the amplitude of the $\pi$ mode, respectively.~\cite{DKZ}  The bare
susceptibilities $\chi_{qq'}^0(\omega)$ are defined, similarly to
Eq.~(\ref{def}), as the Fourier transforms of 
\begin{equation}
\chi_{qq'}^0(t) = -i\theta(t)
\left.\langle[A_{q'}(t),A_q^\dagger(0)]\rangle\right|_{V=J=0}.
\end{equation}
They can be written as components of a $3\times3$ matrix
$\hat{\chi}^0$.  At $T=0$, we have
\begin{equation}
\chi_{qq'}^0(\omega) = \frac{1}{N}\sum_{\bf k}
\left(
\frac{\phi_{q}\phi_{q'}}{\omega-E_{\bf k}-E_{\bf k'}}
-\frac{(-1)^{q+q'}\,\phi_{q}\phi_{q'}}{\omega+E_{\bf k}+E_{\bf k'}}
\right),
\end{equation}
where $E_{\bf k}^2 = \epsilon_{\bf k}^2+\Delta_{\bf k}^2$, ${\bf k' =
Q-k}$, and $\phi_q$ are components of the vector
\begin{equation}
\phi = 
\left(
2^{-1/2}\,p({\bf k,k'}),\ 
g_{\bf k} n({\bf k,k'}),\ 
g_{\bf k} l({\bf k,k'})
\right),
\end{equation}
where $l$, $n$ and $p$ are the BCS coherence factors\cite{Schrieffer}
\begin{eqnarray}
l({\bf k,k'}) &=& u_{\bf k} u_{\bf k'} + v_{\bf k} v_{\bf k'},
\nonumber\\
n({\bf k,k'}) &=& u_{\bf k} u_{\bf k'} - v_{\bf k} v_{\bf k'},
\label{lnp}\\
p({\bf k,k'}) &=& u_{\bf k} v_{\bf k'} + v_{\bf k} u_{\bf k'}.
\nonumber
\end{eqnarray}
The momentum sums are evalulated directly by including a small broadening
$\omega \to \omega + i\Gamma$ in the energy denominators.
Note that, due to strong particle-hole asymmetry implied by our dispersion, 
off-diagonal terms in  $\hat{\chi}^0$ are by no means small.

Within the generalized RPA scheme, the full susceptibilities
$\chi_{qq'}(\omega)$ are given by the matrix RPA equation
\begin{equation}
\hat{\chi}(\omega) = \hat{\chi}^0(\omega) + \hat{\chi}(\omega)\,
\hat{V}\, \hat{\chi}^0(\omega).
\label{mat}
\end{equation}
where $\hat{V}={\rm diag}(-2J,\, V,\, V)$ in the basis (\ref{A_q}).
Recall that both $J$ and $V$ are assumed to be
positive.  We next discuss the behavior of response functions in
various situations.

{\it $J\neq0$, $V=0$}.  This is the situation considered in most
studies.\cite{Lavagna,Levin,Scalapino,Brinckmann} Although the BCS
condensate now mixes all components of $\chi_{qq'}^0(\omega)$, the
full spin susceptibility $\chi_{00}(\omega) = \chi_S(\omega)$ is still
given by the standard RPA expression,
Eq. (\ref{new}).\cite{Brinckmann} This effective decoupling is due to
the fact that the $J$ term in (\ref{interaction}) yields no
interaction in the particle-particle triplet channel.  Despite a
formal similarity with the normal state result, the form of
$\chi_{00}^0(\omega)$ changes dramatically in a d-wave superconducting
state: the Stoner continuum develops a hard gap
\begin{equation}
E_{\rm min}({\bf Q}) = {\rm min}(E_{\bf k}+E_{\bf Q-k}) = 2|\Delta({\bf
k}_{\rm hs})| \approx 2\Delta.
\end{equation}
The imaginary part of 
$\chi_{00}^0(\omega)$ jumps from 0 to a finite value at
$2\Delta({\bf k}_{\rm hs})$.  By Kramers-Kronig relation, this
discontinuity causes a logarithmical divergence of ${\rm Re}
\chi_{00}^0(\omega)$. 
Therefore, for arbitrary $J$, the full $\chi_{00}(\omega)$ has a
resonance pole at $\omega <
2\Delta$.\cite{Levin,Mazin,Scalapino,Brinckmann,Abanov}

In the absence of the triplet-pair coupling $V$, there is no $\pi$
resonance above the continuum.  At the same time, the $\pi$--$\pi$
correlation function contains a resonance below the continuum, which
is nothing but the spin exciton mixing into all other triplet
channels.\cite{Brinckmann,Greiter} Indeed, for $V=0$, the solution of
Eq.~(\ref{mat}) can be written in the following form:
\begin{equation}
\chi_{qq'}(\omega) = \chi_{qq'}^0(\omega) 
-2\chi_{q0}^0(\omega)\, J^{\rm RPA}(\omega)\, \chi_{0q'}^0(\omega),   
\end{equation}
where $J^{\rm RPA}(\omega) = J/[1+2J\,\chi_{00}^0(\omega)]$.  The
spin-exciton pole in the RPA vertex $J^{\rm RPA}(\omega)$ shows up in
all full susceptibilities. 

{\it $J=0$, $V\neq0$}.  In this limit, there is no bound state in the
particle-hole ($q=0$) channel because $J=0$.  The $\pi$ resonance in
the particle-particle channel, which already existed in the normal
state, is pushed to a higher energy when the superconducting gap opens
up.  We stress once again that the lower boundary of the two-hole
continuum $E_{\rm min}$ is produced by fermions with momenta near hot
spots, hence $E_{\rm min} \approx 2\Delta$.  Certainly, an antibound
state in the particle-particle triplet channel is located above
$2\Delta$.  How much above $2\Delta$ depends on the width of the
two-electron continuum and the coupling strength.

The bare $\pi$--$\pi$ response function in a superconducting state is
presented in Fig.~\ref{Fig-bare-pair}(b).  It has two step-like
discontinuities at $E_{\rm min}$ and $E_{\rm max}$, which are seen
in both hole-hole ($\omega<0$) and particle-particle ($\omega>0$)
spectra. The location of the $\pi$ resonance in a superconductor is, however,
not simply given by (\ref{new}), but is a solution of the secular equation
\begin{equation}
\det
\left[
\left(
\begin{array}{ll}
\chi_{11}^0(\omega) & \chi_{12}^0(\omega)\\
\chi_{21}^0(\omega) & \chi_{22}^0(\omega) 
\end{array}\right)
-\frac{1}{V}
\right]
= 0.
\end{equation}
All four matrix elements $\chi^0_{qq'}$ for $q=1,2$ have the same
thresholds as $\chi^0_{\pi\pi}$ in Fig~\ref{Fig-bare-pair}(b).  We
solved this equation numerically and found that the $\pi$ resonance
indeed moves to an energy higher than $2\Delta$.  For $V=40$ meV, the
resonance moves from 22 meV in the normal state
[Fig. \ref{Fig-bare-pair}(a)] to 82 meV in the superconducting state
with $2\Delta = 70$ meV [Fig. \ref{Fig-bare-pair}(b)].  Reasoning
similar to that above implies that the $\pi$ resonance shows up in the
spin channel.
 
\begin{figure}
\begin{center}
\leavevmode
\epsfxsize 3.25in
\epsffile{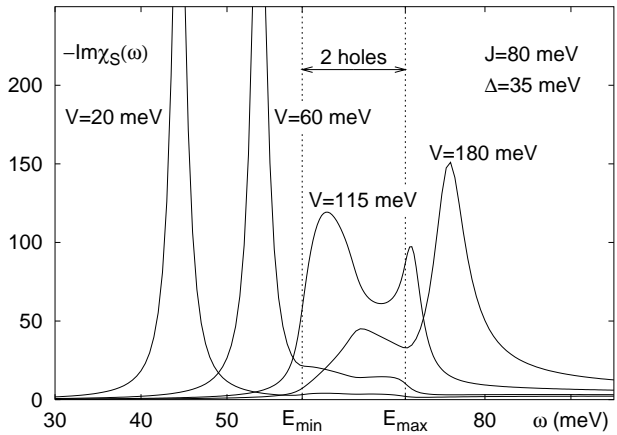}
\end{center}
\caption{Spin susceptibility [$-{\rm Im}\chi_S(\omega)$] 
in the superconducting state at ${\bf Q}$.  Values of $V$ are shown
near each curve.  $E_{\rm min}$ and $E_{\rm max}$ are boundaries of
the two-hole continuum.  The chosen values for $J$ and $\Delta$ are
similar to those measured by neutron and ARPES, respectively.}
\label{Fig-dressed}
\end{figure}

{\it $J\neq0$, $V\neq0$}.  The general case interpolates between the
two limits.  Fig. ~\ref{Fig-dressed} presents our results for the spin
susceptibility at various ratios of the coupling strengths $V/J$.
Interestingly enough, for comparable couplings, we found that the
response functions contain {\it two} peaks at different energies.
One, above $E_{\rm max}$, is the $\pi$ resonance, the other, below
$E_{\rm min}$, is the bound state in the particle-hole channel.  As
$V$ is increased, the high-energy resonance pulls away from the upper
edge of the two-hole continuum and stregthens.  At the same time, the
low-energy resonance approaches the lower edge and enters into the
two-electron continuum, where it broadens and finally becomes
invisible.  In the opposite limit, as $V$ gets progressively smaller,
the $\pi$ resonance weakens, merges with the outer edge of the
continuum and disappears.

{\em Discussion.}  Direct comparison\cite{Zasadzinsky} of neutron and
ARPES data for \bscco\/ shows that the neutron resonance is located
well below $2\Delta$.  We believe that the same holds true for \ybco.
Because the lower continuum edge $E_{\rm min}$ is just below
$2\Delta$, the resonance almost certainly occurs below this edge.
This point can be verified by analysing ARPES data at the hot
spots. In a slightly overdoped \bscco\/ sample with $T_c=87$ K, the
gap value at the hot spots is $32\pm3$ meV,\cite{Ding} and $E_{\rm
min} \geq 58$ meV.  The neutron resonance at ${\bf Q}=(\pi/a,\pi/a)$
in the overdoped \bscco\/ with similar $T_c=83$ K is observed at $E =
38$ meV,\cite{peak-discovery-bscco,He-cond-mat} i.e., well below
$E_{\rm min}$.  We thus conclude that the neutron resonance is an
exciton-like bound state in the particle-hole channel, not an
antibound state, such as the $\pi$ resonance.

It has been previously remarked that particle-hole
mixing can mask the origin of the neutron
resonance:\cite{Brinckmann,DKZ,Greiter} a spin resonance mixes into
triplet-pair channels and vice versa.  We rely on a continuity
argument to make the unambiguous identification of the experimental
neutron peak, which is observed below the continuum edge $E_{\rm min}
\approx 2\Delta$, as expected for $V \ll J$.  As the pair-triplet
coupling $V$ is turned off, the low-energy peak stays below the edge
and continuously evolves into the usual RPA spin resonance at
$V=0$.  On the other hand, there is no collective mode below $E_{\rm
min}$ in the limit $V \gg J$, in which the $\pi$ resonance can be
defined unambiguously.  Moving the high-energy $\pi$ resonance across the
continuum involves a discontinuous change.  For this reason, the
neutron resonance should not be associated with the $\pi$ resonance.

We feel that a more realistic analysis will not change this
general conclusion because our argument does not rely on detailed
dynamics of the resonances, but is based on measured kinematics of
low-energy fermion excitations and on general properties of bound
states.  If one wishes to find the $\pi$ resonance in neutron scattering,
the search should be confined to energies above the two-hole continuum.
As its true upper edge is not known experimentally, we can
only suggest that this energy exceeds $2\Delta$.  On the other hand,
because of the large incoherent parts of the spectral function observed
in ARPES data for energies beyond $\Delta$, there may be no true upper edge
to the continuum,\cite{BA} and it is quite possible that the $\pi$ resonance
will be strongly damped, or perhaps even absent, in neutron data.

The authors thank Ar. Abanov, J. C. Campuzano, E. Demler, B. Jank\'o,
and J. F. Zasadzinski for helpful discussions.  A. C. and O. T. are
grateful to ANL staff for hospitality during their stay at Argonne,
and M. N. to the staff at the ITP.
This work was supported by the DOE Office of Science under contract
W-31-109-ENG-38 (Argonne), by the DOE Grant No. DE-FG02-90ER40542
(Princeton) and by the NSF grants DMR--9979749 (Wisconsin)
and PHY94-07194 (ITP).

\end{document}